\documentclass[a4paper,12pt]{article}
\usepackage[utf8]{inputenc}
\usepackage{cancel}
\usepackage{ulem}
\usepackage{amsfonts}
\usepackage{amssymb}
\usepackage{graphicx}
\usepackage{amsmath}
\usepackage[table,xcdraw]{xcolor}
\usepackage{tikz}
\usetikzlibrary{decorations.markings}
\usepackage{enumerate}
\usepackage{mathtools}
\usepackage{subfloat}
\usepackage{subfig}
\usepackage{color}
\usepackage{tikz}\usetikzlibrary{calc}
\usepackage{float}
\usepackage{here}
\usepackage{cite}
\usepackage{mathrsfs}
\usepackage{float,epsfig}
\usepackage{dcolumn}% Align table columns on decimal point

\usepackage{graphicx}% Include figure files
\usepackage{bm}% bold math
\usepackage{amsmath,amssymb,amsthm}
\usepackage[colorlinks=true,linkcolor=blue,citecolor=red]{hyperref}
\newcommand{\be}{\begin{equation}}
\newcommand{\ee}{\end{equation}}
\newcommand{\bea}{\setlength\arraycolsep{2pt} \begin{eqnarray}}
\newcommand{\eea}{\end{eqnarray}}

\def\0{{\sst{(0)}}}
\def\1{{\sst{(1)}}}
\def\2{{\sst{(2)}}}
\def\3{{\sst{(3)}}}
\def\4{{\sst{(4)}}}
\def\5{{\sst{(5)}}}
\def\6{{\sst{(6)}}}
\def\7{{\sst{(7)}}}
\def\8{{\sst{(8)}}}
\def\sst#1{{\scriptscriptstyle #1}}

\definecolor{lime}{HTML}{A6CE39}
\newcommand{\orcidicon}{%
    \begin{tikzpicture}
    \draw[lime, fill=lime] (0,0)
        circle [radius=0.16]
        node[white] {{\fontfamily{qag}\selectfont \tiny ID}};
    \draw[white, fill=white] (-0.0625,0.095)
        circle [radius=0.007];
    \end{tikzpicture}   \hspace{-2mm}
}
\newcommand\orcidAdil{{\href{https://orcid.org/0000-0001-7623-5541}{\orcidicon}}}
\newcommand\orcidHasan{{\href{https://orcid.org/0000-0001-7408-0910}{\orcidicon}}}

\makeatletter \@addtoreset{equation}{section}

\usepackage{multirow}
\begin{document}
%	\begin{flushright}
%	ksdjflk kjsdlfkj
%	\end{flushright}
%

\title{\normalsize
%\phantom{fff}
%\vspace{-3cm}
%\begin{flushright}
%FISPAC-TH/27/2020\\
%UQBAR-TH/2020-314
%\end{flushright}
%\vspace{2cm}
%%%
%%
%%
{\bf \Large	  Optical Shadows  of Rotating Bardeen AdS Black Holes  }}
\author{ \small   A. Belhaj\orcidAdil\!\! $^{1}$\footnote{a-belhaj@um5r.ac.ma},  H. Belmahi$^{1}$\footnote{hajar\_belmahi@um5.ac.ma},  M. Benali\orcidAdil\!\!$^{1}$\footnote{mohamed\_benali4@um5.ac.ma}, H. El Moumni\orcidHasan\!\!$^{2}$\thanks{hasan.elmoumni@edu.uca.ma},
	M. A. Essebani$^{3}$,  M. B. Sedra$^{3,4}$\footnote{ Authors in alphabetical order.}
	\hspace*{-8pt} \\
	%EndAName
	{\small $^1$ D\'{e}partement de Physique, Equipe des Sciences de la mati\`ere et du rayonnement, ESMaR}\\
{\small   Facult\'e des Sciences, Universit\'e Mohammed V de Rabat, Rabat,  Morocco} \\
	{\small $^{2}$  EPTHE, D\'{e}partement de Physique, Facult\'e des Sciences,   Universit\'e Ibn Zohr, Agadir, Morocco} \\
	{\small $^{3}$  D\'{e}partement de Physique,  Laboratoire de physique des Mat\'eriaux et Subatomique, LPMS}\\   {\small Facult\'{e}
		des Sciences, Universit\'{e} Ibn Tofail, K\'{e}nitra,
		Morocco } \\ {\small  $^4$ Moulay Ismail University, FSTE, LSTI, BP-509 Boutalamine,    Errachidia 52000,  Morocco    }
} \maketitle

\begin{abstract}
		{\noindent}
Using the Hamilton-Jacobi method, we study certain optical shadows of rotating Bardeen-AdS black holes in the presence of internal and external field contributions.  Precisely,  we examine the shadow using one-dimensional real curves. By establishing the equations of motion of such AdS black holes without external dark sectors, we first analyze the shadow configurations in terms of a reduced moduli space involving only internal parameter contributions including the charge of the nonlinear electrodynamics.  Among others, we find that this parameter serves as a geometric quantity controlling the shadow shape.   Then,  we  graphically discuss  the associated astronomical observables.   Enlarging the moduli space, we analyze the behaviors of the quintessential black hole solutions in terms of the involved parameters controlling the dark field sector.   Concretely, we observe that the dark energy not only affects the size but also deforms the shadow shape.  Finally, we   provide  a possible link 
with observations from Event Horizon Telescope    by  showing  certain constraints on the involved  moduli space  in
the light of the M87$^*$  picture.
	\end{abstract}
\newpage
\newpage
\section{Introduction}
Black hole physics   has received  immense attention since the  thermodynamic   activities including the Hawking-Page phase transitions elaborations.    For  extended phase space scenario, the black hole thermodynamics  has  been studied by considering  the cosmological constant as  a dynamical quantity identified with the pressure. In  such  a picture,  the charged AdS black holes  undergo  phase  transitions similar to ones  appearing in  van der Waals fluids.\\
Within the general relativity, the Penrose and Hawking theorems stipulate the existence of a physical singularity at  the black hole center \cite{x19,x20} . Nonetheless, there are  various  methods to get  the physics  around this singularity.  Certain singularity-free black hole solutions can be obtained. One of such interesting classes of these solutions   has been obtained  from  gravity  models  in the presence of non-linear electrodynamics couplings\cite{x21, x22}. The particularity of such solutions  concerns  smooth  spacetime   being  free from a singularity because the mass parameter is only a consequence of a non-linear electrodynamics source.   These regular solutions can be considered as   gravitational field  contributions   with non-linear electric or magnetic monopoles.
This formalism  concerns  two special classes, namely the Bardeen
and the  Hayward solutions \cite{x23, x24}. Such black holes in Anti-de-Sitter (AdS)  spacetime exhibit  a. phase portrait  as  the one of van der Waals fluids\cite{x25, x26}.\\
The recent astronomical observations  have shown that the universe is undergoing accelerating expansions,  which   has  provided   a negative pressure state \cite{x40,x41}.  It has been suggested that  this   feature associated with  accelerated expansions is a direct consequence of the presence of    field  configurations   known   by  dark energy (DE). Although the cosmological constant, the quintessence is one of the different models for dark  sectors\cite{x74,x75}.  It is recalled that the  cosmic source for inflation involves  the equation of state $p_q = \omega \rho_q$ with ($-1 < \omega < -1/3$), and $\omega = -2/3$ corresponds to the quintessential  dark energy regime.  Such a DE   involves  the form $p_q=-\frac{\alpha}{2}\frac{3\omega}{r^{3(\omega+1)}}$,   where $\alpha$ a relevant parameter.  Phase transitions in  quintessential black holes are intensively investigated from different angles. In the presence of  such a dark sector,  the thermodynamics of charged black holes  and  the regular ones     has been  dealt with  \cite{x77,x78,x79,x80,x801,x802,x803}.  Moreover,  the geothermodynamics for different regular black holes  has been studied in \cite{x65,B5,J1}, while  the  case of the charged AdS black holes surrounded by such  quintessence field contributions  can be found in \cite{x83,x84}.\\
Black hole optical physics has received remarkable  interest using different roads.  These activities have been supported   by  considerable  efforts of Event Horizon Telescope collaborations providing  an  interesting  picture  of supermassive black holes at the center of galaxy $M87$ \cite{3,4}. Several  works have been elaborated dealing with   such optical behaviors     by approaching   the  shadow geometry and the deflection angle \cite{B2,B6}. In particular, the black hole shadow behaviors of various  black holes have been examined\cite{B13, B10}.   Using  the null geodesic  relations, the elaboration of the shadow cast has been   investigated  by  studying the associated geometrical  behaviors. 
  These behaviors have been   controlled  by certain  astronomical observables.  The geometrical quantities  have been extensively studied  for different black hole backgrounds  and gravity models \cite{R1}.  They provide  certain physical information  about   the associated  space-time geometries.  An examination shows that two  observables have been adopted  needed to control  the size and  the shape  geometric deformations of one dimensional real manifolds.  Using such activities, the  shadow  distortion behaviors   of Einstein-Maxwell-Dilaton-Axion black holes  have  been examined \cite{R1}.   In the absence of the rotating parameter,  it has been   demonstrated  that the shadows  involve  a circular  picture  where  the associated  size can be  deformed  by the mass and other parameters  describing  the charge and  the dark   field intensity \cite{B1}.  For rotating black hole solutions,   however,  this  geometric configuration     has been  distorted    by  providing new geometries.  These involve   D-shape  and other pictures including  the cardioid  geometries\cite{BC}.  Moreover, it has been observed  that  the shadow  aspects could  depend  also on  geometric and stringy  parameters.   Indeed,   the brane number  and cosmological scale     have been implemented  where non-trivial geometries have been appeared \cite{B11,B12}. 
 \\
In this work, we aim to investigate the optical behaviors of Bardeen AdS black holes by examining  the  corresponding  shadows using  one dimensional  real curves.  By   establishing the  equations of motion of  such AdS black holes without external dark sectors, we   first   discuss  the shadow geometries in terms of a reduced  moduli space  involving  only  internal  parameter contributions including  the charge of the nonlinear electrodynamics.  Among others, we   find that this parameter   controls  the shadow  shape.   Then,  we  analyze  graphically the  corresponding astronomical observables.   Enlarging  the moduli space,  we     approach   the shadow  optical behaviors  of quintessential  black hole solutions in terms of the involved parameters.  Precisely,  the combination of the rotating and  the DE parameters of  the Bardeen AdS black holes shows that the geometry  of  the shadows has been  affected by the DE  sector contributions.  For non-rotating case, the associated  parameter  can  control the size.  However,  it   contributes also  to the  shadow distortions  for rotating cases. 
 Finally, we   present   a possible connection
with observations from  the Event Horizon Telescope (EHT)    by giving   certain constraints on the involved  black hole parameters in
the light of the M87$^*$  picture.\\
 This work is organized as follows. In section 2,  we present a concise review of rotating Bardeen-AdS black holes in four dimensions. In section 3,   we investigate the shadow behaviors of rotating Bardeen-AdS black holes without external dark field contributions. In section 4, we study the quintessential Bardeen AdS black holes.
  In section 5, we  show  a possible connection with M87$^*$  observations. 
 The last section concerns concluding remarks and open questions.
\section{Rotating Bardeen-AdS black holes}
 In this section, we  investigate the optical aspect of the  charged and rotating  Bardeen black holes in four dimensions with AdS geometries.
Concretely, we  examine  the shadow geometrical pictures in terms of many parameters
including the cosmological constant $\Lambda$.  It is interesting to note  that two solutions can arise depending on such a constant.  For  $\Lambda>0$, the
model will be called Bardeen de Sitter (Bradeen-dS). However, $\Lambda<0$ provides a solution 
referred to as  Bardeen  Anti de Sitter (Bardeen-AdS) which will be  studied in certain
details through this work by assuming that similar computations could done for the first solution.  Moreover,   it    has been known from the  previous works that the black hole shadow with negative cosmological constant is more large than that   the ones corresponding to  the positive one \cite{B13}.\\
Before investigating the optical behaviors,  we  give first  a concise review  on rotating Bardeen-AdS black holes in four dimensions.    Such solutions have been extensively studied\cite{T1}.   In  Boyer-Lindquist coordinates, the line element, associated with  solutions without  external contributions describing the dark sector,    reads as
\begin{equation}
ds^{2}=-\frac{\Delta_{r}}{\Sigma}\left(dt-\frac{a sin^{2}\theta}{\Xi}d\phi\right)^{2}+\frac{\Sigma}{\Delta_{r}}dr^{2}+\frac{\Sigma}{\Delta_{\theta}}d\theta^{2}+\frac{\Delta_{\theta} sin^{2}\theta}{\Sigma}\left(a dt - \frac{r^{2}+a^{2}}{\Xi}d\phi\right)^{2}.
\end{equation}
The involved reduced terms are  expressed as follows
\begin{equation}
\begin{array}{ll}
\Delta_{r}= \left(r^{2}+a^{2}\right)\left(1+\frac{r^{2}}{\ell^{2}}\right)-2m\left(\frac{r^{2}}{r^{2}+g^{2}}\right)^{\frac{3}{2}}r, \hspace{2.9cm} \Delta_{\theta}= 1-\frac{a^{2}}{\ell^{2}}cos^{2}\theta, \\ [6px]
 \hspace{ .25cm}\Xi = 1-\frac{a^{2}}{\ell^{2}}, \hspace{ 8.5cm} \Sigma = r^{2}+a^{2} cos^{2}\theta.
\end{array}               \end{equation}
Here,  $m$ and $a$ represent the mass and the rotating parameter of the black hole, respectively. The quantity $g$ is the charge of the
nonlinear electrodynamics.   The parameter $\ell$  denotes  the AdS radius being  related to the pressure $P$  via $
P=-\frac{\Lambda}{8\pi}=\frac{3}{8\pi \ell^{2}}$. These solutions can be extended by implementing external dark field contributions\cite{x74}.   A particular emphasis has been put on DE  using  the quintessence field contributions.   In this way,   the above delta  term can be  generalized as follows
\begin{equation}
\label{q1}
\Delta_{r}^\omega= \left(r^{2}+a^{2}\right)\left(1+\frac{r^{2}}{\ell^{2}}\right)-2m\left(\frac{r^{2}}{r^{2}+g^{2}}\right)^{\frac{3}{2}}r-\frac{\alpha}{r^{3\omega+1}},
\end{equation}
where $\alpha$  denotes the  intensity of quintessence and $\omega$ is the quintessential state parameter.    The optical  behaviors of these solutions will be elaborated. First, we consider  solutions without dark field contributions. Then, we analyze the effect of such an external dark field.  This study will be done in terms of a moduli space  coordinated by  the relevant parameters  including the ones  associated with dark   field sectors.

\section{Shadow properties  of  rotating   Bardeen AdS black holes}
In this section, we   approach    the optical  aspects  of   the  rotating    Bardeen-AdS black holes without external dark field contributions.    In particular, we consider the   investigation of   the shadow  optical  behaviors   in terms of  the involved   internal parameters including  the AdS radius. This study will be done for particular regions of the  moduli space.
\subsection{Optical shadows}
\begin{figure}[ht!]
		\begin{center}
		\centering
			\begin{tabbing}
			\centering
			\hspace{9.4cm}\=\kill
			\includegraphics[scale=.45]{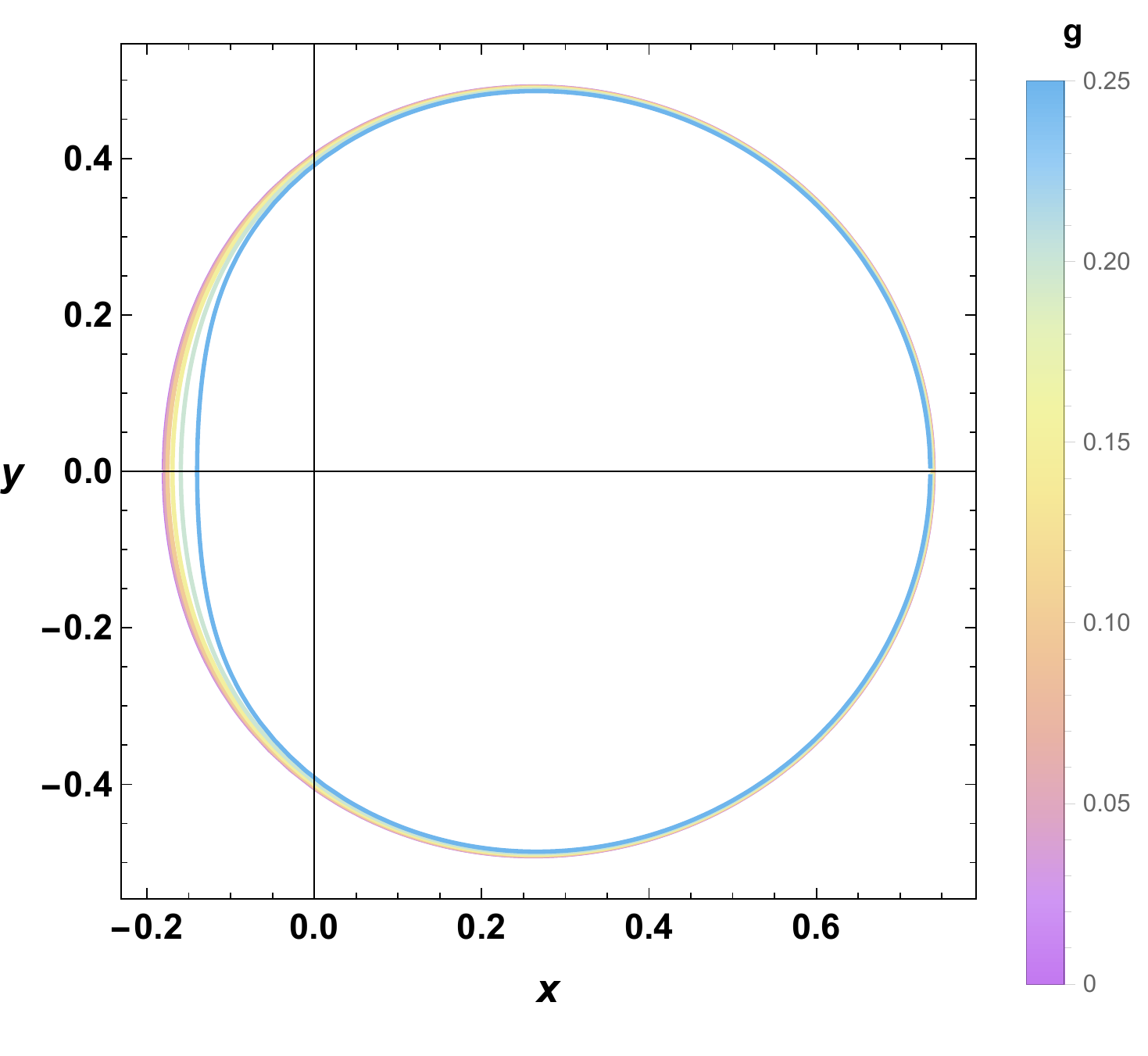} \>
			\includegraphics[scale=.45]{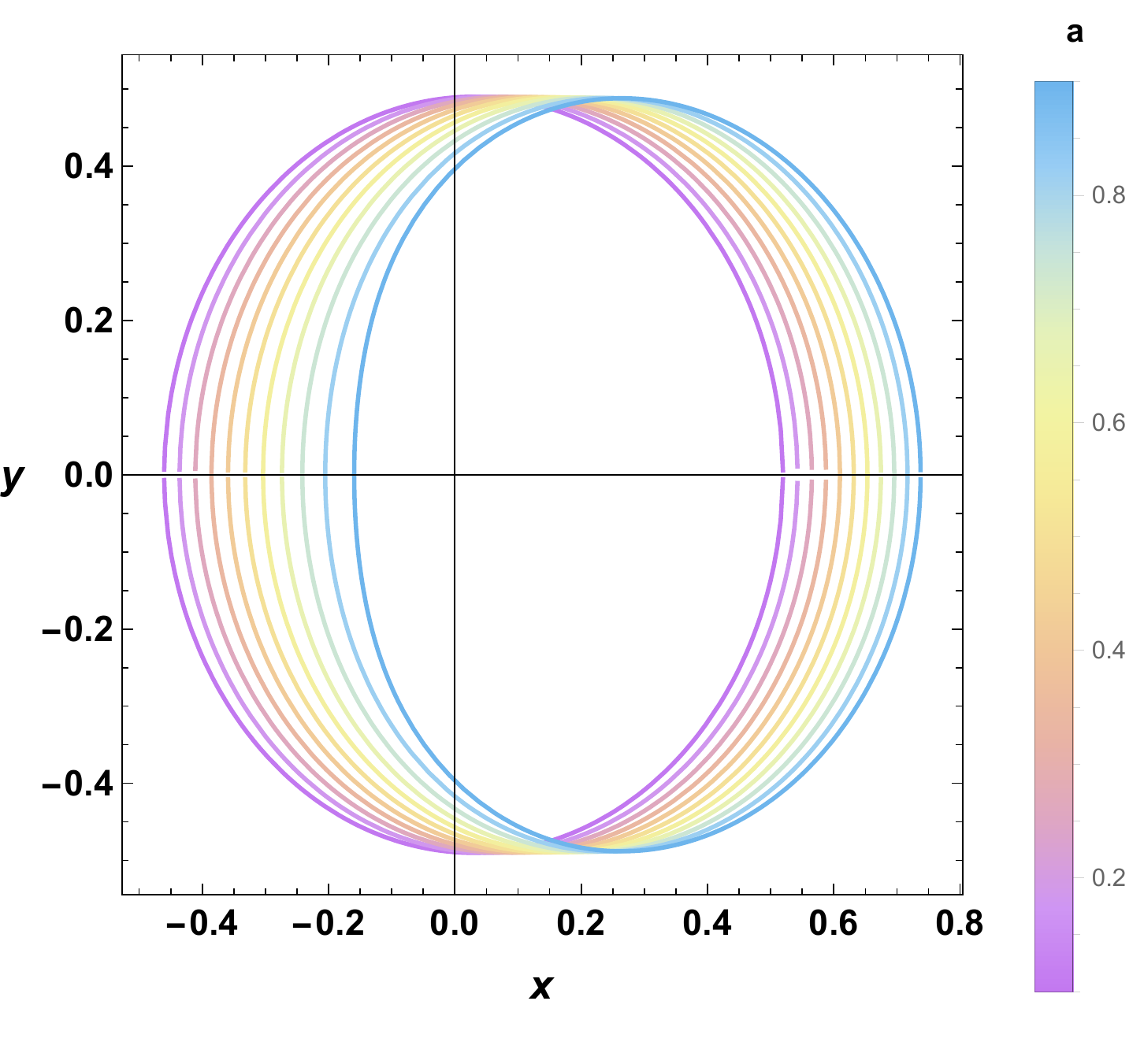} \\
		   \end{tabbing}
		    \vspace{- 1 cm}
\caption{{\it \footnotesize   Optical shadow aspects of  rotating    Bardeen-AdS black holes. Right panel: behavior  associated with $a = 0.9$  for different  values $g$. Left panel: behavior corresponds to  $g = 0.2$   for different values  $a$}.}
\label{shfa1}
\end{center}
\end{figure}

To approach the shadow optical  aspects,  certain relations are needed.  Indeed, we first   write down the massless particle equations of motion by exploiting the Hamilton-Jacobi method\cite{A2}. For a photon in the black hole spacetime, one  should  use
\begin{equation}
\frac{\partial S}{\partial \tau }=-\frac{1}{2}g^{ij}p_{i}p_{j},
\end{equation}%
where $S$ and $\tau $  are the Jacobi action and the affine
parameter respectively, along the geodesics. In the spherically symmetric spacetime,  the Jacobi action $S$    takes the following form
\begin{equation}
\label{ks2}
S=-Et+L\phi+S_r(r)+S_\theta(\theta). 
\end{equation}
In this action,   $E$ and $L$   represent   the conserved  total energy and  the  conserved angular momentum of the  massless particle,  respectively.   In four dimensions,  they  are written  in terms of the four-momentum  $p_\mu$  as follows    $E=-p_t$ and $L=p_\phi$   which are geodesic
constants of motion. It is denoted that  $S_\theta(\theta)$ and  $S_r(r)$ and   are functions which   involve only $\theta$ and $r$   spatial variables, respectively.
The relevant   null geodesic equations  can be obtained  using a separation method  similar to the one exploited in  the Carter mechanism \cite{A1,A2}.  To approach  the  black hole shadow  configurations,    two  quantities called  impact parameters     are needed.  In terms of the above conserve quantities, they  are given  by the following relations 
\begin{equation}
\label{xe}
 \xi=\frac{L}{E}, \hspace{1.5cm}\eta=\frac{\mathcal{K}}{E^2},
\end{equation}
where $\mathcal{K}$   is a separable constant shearing aspects as the one   proposed in  \cite{A1}.   According to  the shadow activities developed, in many works,   one can  get the  null geodesic relations. Indeed, they are given by 
\begin{align}
\Sigma \frac{d  \, t}{d \tau}& =  E \left[ \frac{\left(r^2 +a^2 \right)\left[ \left(r^2 +a^2 \right)- a \xi \Xi  \right] }{\Delta_r} +  \frac{ a \left(\xi \Xi - a  \sin^2 \theta \right) }{ \Delta_\theta} \right]\\
\Sigma  \frac{d  \, r}{d \tau} &=\sqrt{\mathcal{R}(r})\\
 \Sigma\frac{d  \, \theta}{d \tau} & =\sqrt{\Theta(\theta)}\\
\Sigma \frac{d  \, \phi}{d \tau} &= E \;\Xi \left[ \frac{a(\left(r^2 +a^2 \right) -a \xi \Xi)}{\Delta_r} + \frac{ \xi\Xi-a \sin^2 \theta}{\sin^2 \theta \Delta_\theta}  \right].
\end{align}
 In this  model,    $\mathcal{R}(r)$ and ${\Theta}(\theta)$    are expressed as follows
 \begin{align}
\mathcal{R}(r)&=E^2\left[ \left[ \left( r^2 +a^2 \right)  -a \xi \Xi  \right]^2- \Delta_r\eta   \right]\\
 \Theta(\theta)&= E^2 \left[ \eta \Delta_\theta  - \csc^2 \theta\left( a\sin^2 \theta - \xi \Xi\right )^2 \right]
\end{align}
indicating  the radial and  the polar motion, respectively.   Roughly, the  black hole shadow  shapes  can be elaborated   from   unstable circular orbits.  Precisely,  this can be obtained by using the following  conditions
\begin{equation}\label{xx1}
\mathcal{R}(r)\Big|_{r=r_0}=\frac{d\mathcal{R}(r)}{d r}\Big|_{r=r_0}=0.
\end{equation}
Here, $r_0$    denotes   the circular orbit radius of the photon massless particle\cite{B1,A2}.   Taking  $ \Theta(\theta)>0$ for $0\leqslant\theta\leqslant2\pi$, the  above impact parameters   can be obtained by  solving Eqs.\eqref{xx1}.  Computations provide
\begin{align}
& \eta = \frac{16 r^2 \Delta_r}{\Delta _r^{\prime\,2}},\\
& \xi=\frac{\left(r^2+ a^2\right) \Delta_r^\prime-4r \Delta_r  }{a\Xi \Delta_r^\prime} \bigg\vert_{r=r_0},
\end{align}
where the spatial  derivation  $\Delta_r^\prime=\frac{\partial \Delta_r}{\partial r}$ has been used.  A close examination shows that in non-trivial backgrounds the  shadow computations need certain relevant considerations.  The cosmological constat $\Lambda$,  for instance,  brings a visible way  to approach such considerations.  In fact,   one should fix  the position ($r_{ob}$\,,$\theta_{ob}$) of the observer  using  Boyer-Lindquit  system coordinates\cite{D3,D2}. It is recalled that   $r_{ob}$  and  $\theta_{ob}$  represent  the radial and the angular coordinates, respectively.  The calculations will be based on appropriate assumptions.  Placing  the observer  in the  outer communication domain ($\Delta_r >0$), and  assuming that  the light ray trajectories    are  ejected   from the  position ($r_{ob}$\,,$\theta_{ob}$) to past\cite{D1,D2},   the  observer  position can be approached  by suing 
 the orthogonal tetrads ($e_0$,\,$e_1$,\,$e_2$,\,$e_3$).  In terms of the metric black hole data, they are given by 
\begin{eqnarray}
\label{e_0}
e_0 & = & \frac{ (r^2+a^2)\partial_t+a\Xi\partial_\phi}{\sqrt{\Delta_r \Sigma}} \bigg\vert_{(r_{ob}\,,\theta_{ob})}\qquad 
e_1  =  \frac{\sqrt{\Delta_\theta}}{\sqrt{\Sigma}} \partial_\theta\bigg\vert_{(r_{ob}\,,\theta_{ob})}\\
\label{e_2}
e_2 & = &  -\frac{ a\sin^2{\theta}\partial_t+\Xi\partial_\phi}{\sqrt{\Delta_r \Sigma}\sin{\theta}} \bigg\vert_{(r_{ob}\,,\theta_{ob})}\qquad 
e_3  = -\frac{\sqrt{\Delta_r}}{\sqrt{\Sigma}} \partial_r\bigg\vert_{(r_{ob}\,,\theta_{ob})}
\end{eqnarray}
where   the timelike vector $e_0$  denotes    the four-velocity of the  observer.     However,   $e_3$  is identified with     the spatial direction  towards  the black hole hole center.  The directions $e_0\pm e_3$  represent  tangent to the principal null congruences.  In this vector representation,  the light rays  can be  parametrized by $
\lambda(s)=(r(s),\theta(s),\phi(s),t(s))$.
One can also  consider the tangent vector  $\dot{\lambda}$  which  can be decomposed  as follows
\begin{equation}
\label{lambdatt}
\dot{\lambda}=\alpha(-e_0+\sin\rho\cos\delta\,e_1+\sin\rho\sin\delta\,e_2+\cos\rho\,e_3),
\end{equation}
where $\alpha$  is a scalar factor.  $\rho$ and $\delta$  are  the celestial coordinates  \cite{D1}.   After  appropriate computations, one  can obtain 
\begin{equation}
\label{alpha}
\alpha=g(\dot{\lambda},e_0)=\frac{1}{\sqrt{\Delta_r\Sigma}}(aL\Xi-(r^2+a^2)E) \bigg\vert_{(r_{ob}\,,\theta_{ob})}.
\end{equation}
The examination of  such black hole  optical shadows needs the implementation of  all  parameters including the geometric ones.  Concretely, a   reduced moduli space will control the associated  optical shadows.  Following \cite{D1,D2,D3}, the boundary of   the geometrical optical  shadows  can be  approached  by the help of two cartesian coordinates
\begin{eqnarray}
\label{x}
x & = & -2 \tan{(\frac{\rho}{2})}\sin{\delta}\\
\label{y}
y & = & -2 \tan{(\frac{\rho}{2})}\cos{\delta}
\end{eqnarray}
where the   celestial coordinates $\rho$ and $\delta$  are given  in terms of $\xi$ and $\eta$ as follows
\begin{eqnarray}
\label{rho}
\sin{\rho}&=&\frac{\pm\sqrt{\Delta_r\eta}}{((r^2+a^2)-a\xi\Xi)}\bigg\vert_{(r_{ob}\,,\theta_{ob})}\\
\label{psi}
\sin{\delta}&=&\frac{\sqrt{\Delta_r}\sin\theta}{\sqrt{\Delta_\theta}\sin\rho}\left(\frac{\Xi(a-\Xi\csc^2{\theta}\xi)}{a\Xi\xi-(r^2+a^2)}\right)\bigg\vert_{(r_{ob}\,,\theta_{ob})}.
\end{eqnarray}
A close  inspection  reveals  that the shadow geometrical  behaviors  can depend on the involved parameters describing  the   rotating black holes. In these solutions,  the charge of the nonlinear electrodynamics will be considered as a relevant one.   The  corresponding  shadow properties  are represented in  the   $(x,y)$ plane by  using  $x$ and $y$  explicit expressions.   For the present study,   certain  parameter values and positions  have   been fixed.   Concretely,     the observer is  located at $r_{ob}=50$ and $\theta_{ob}=\frac{\pi}{2}$. It is denoted $\Lambda=-10^{-4}$ and  $m=1$ have been  taken. The associated  behaviors  are plotted  in  Fig.(\ref{shfa1}). Concretely, we  illustrate  the shadow   aspects  by varying  the  two parameters $g$ and $a$.  It  has been remarked   that the shadow for a  fixed value of $a=0.9$ is not a perfect circle. Taking  $a=0.9$,   the D-shape   geometrical  configurations appear by increasing $g$.  Considering  $g=0.2$, the circles have been distorted and deviated from the centre by increasing  $a$. Such a  D-shape appears for large values of the rotating parameter. In these rotating AdS black holes,   it has been shown that the D-shape geometry is linked only to the rotating parameter $a$.  In the present situation, however, the D-shape configuration  is linked  not only to   the  rotating parameter $a$ but also to the charge of the nonlinear electrodynamics $g$.   Concretely,  the D-shape geometry is controlled by  the parameters  $a$ and $g$.  It  has been observed  that  the  parameter $g$ affects the space-time configuration of the black hole and the shadow observation form.

 \subsection{Distorted optical shadows}
\begin{figure}[ht!]
		\begin{center}
		\centering
			\begin{tabbing}
			\centering
			\hspace{9.4cm}\=\kill
			\includegraphics[scale=.45]{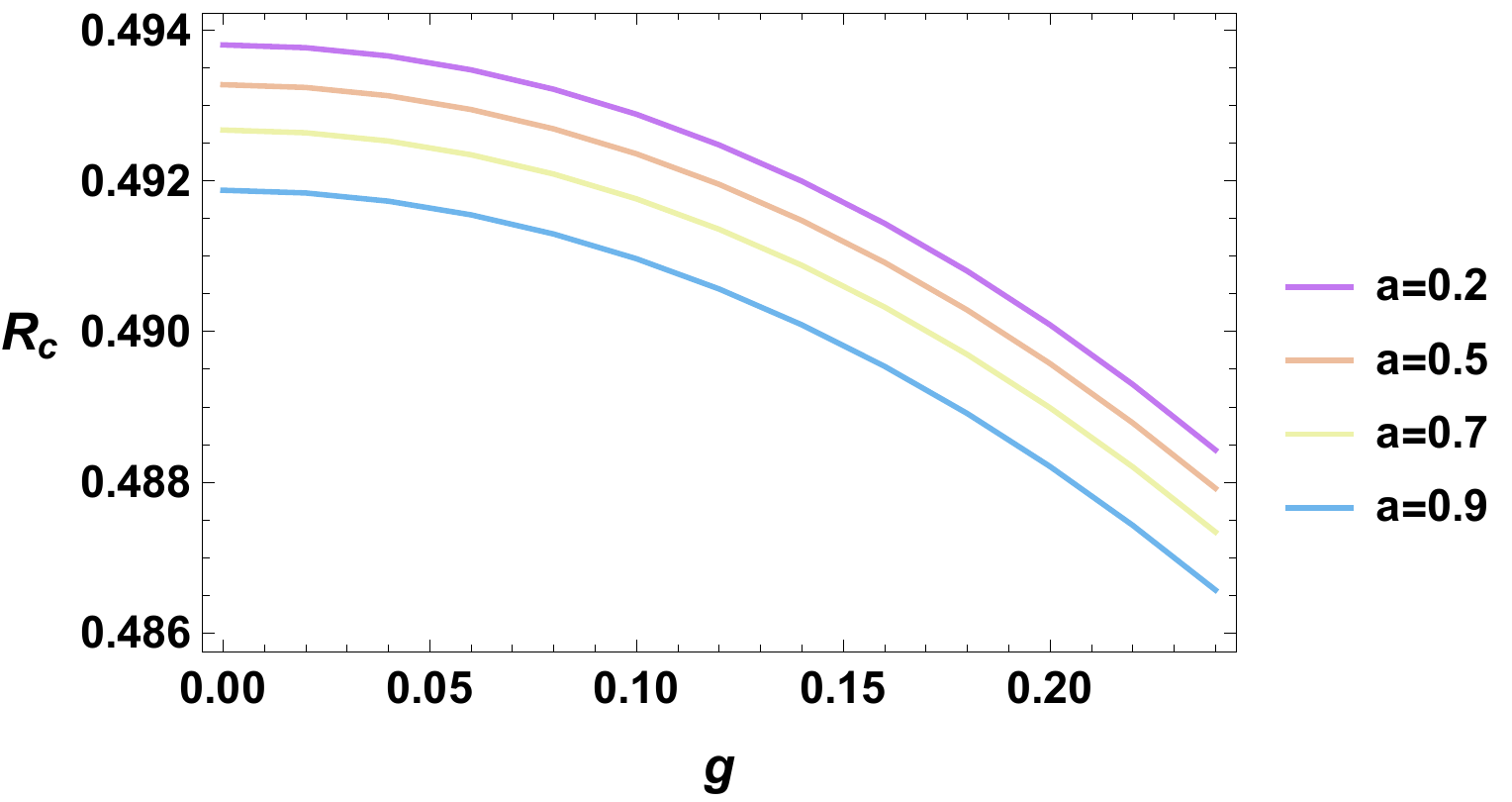} \>
			\includegraphics[scale=.45]{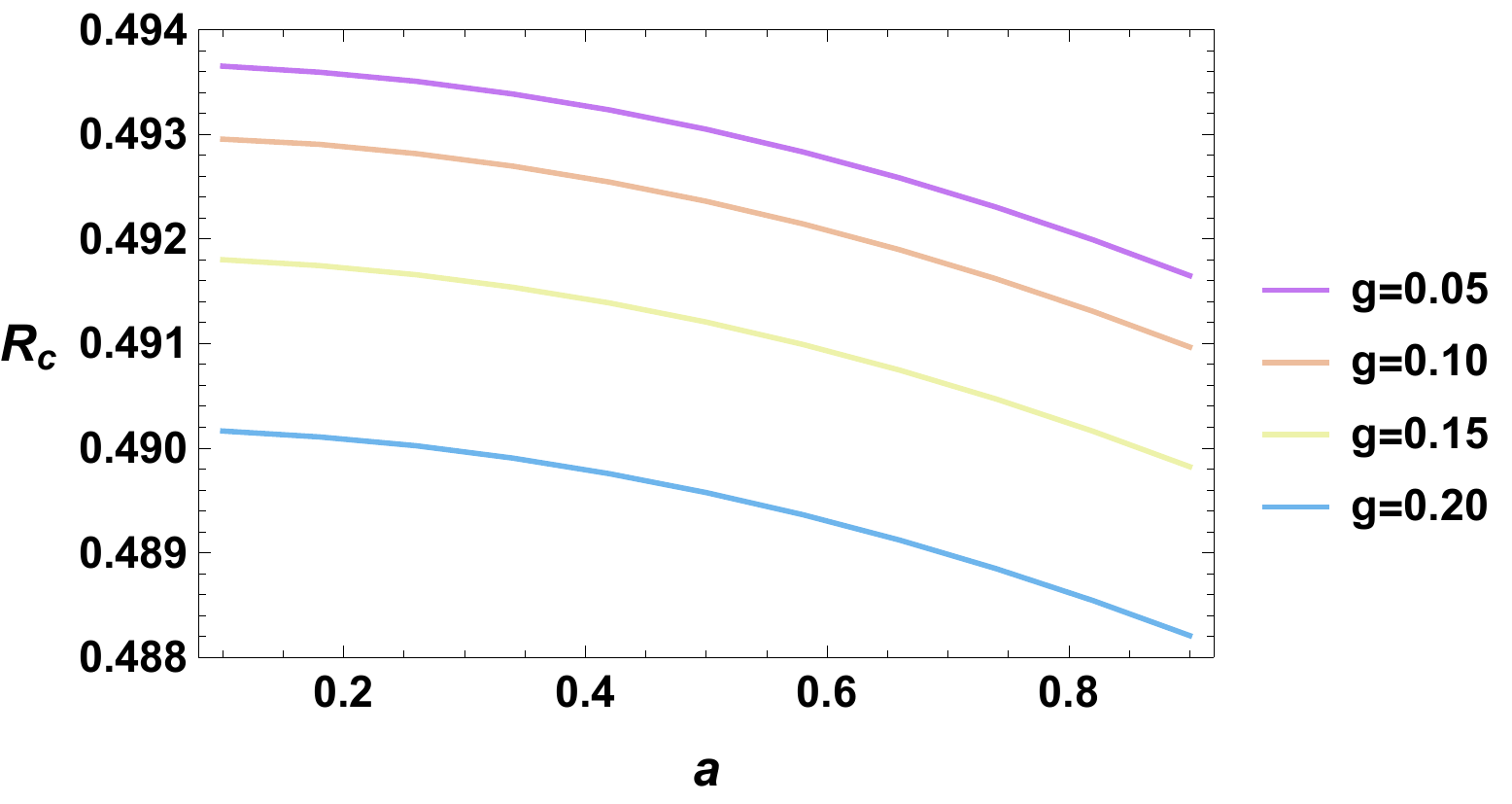} \\
			\includegraphics[scale=.45]{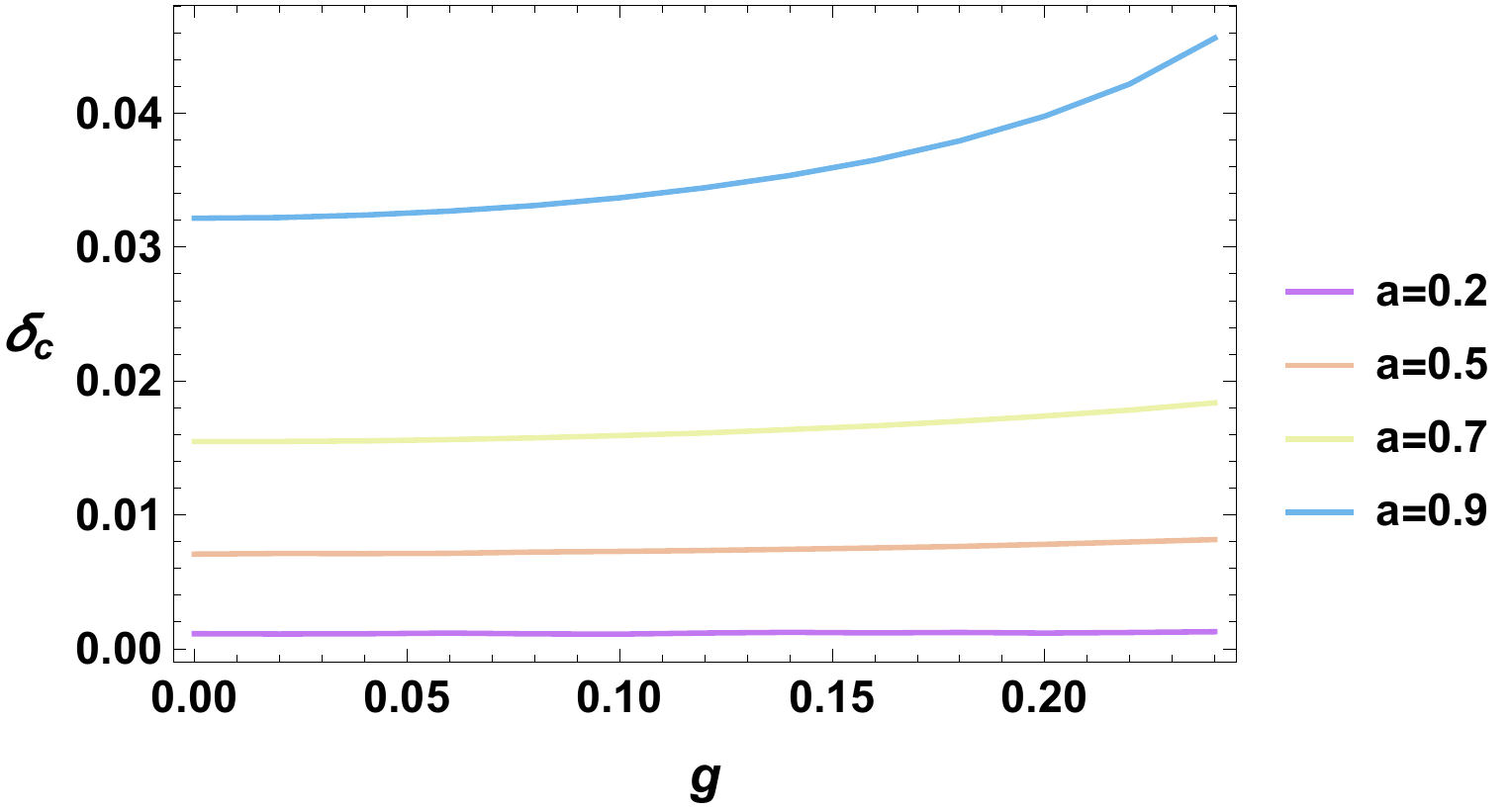} \>
			\includegraphics[scale=.45]{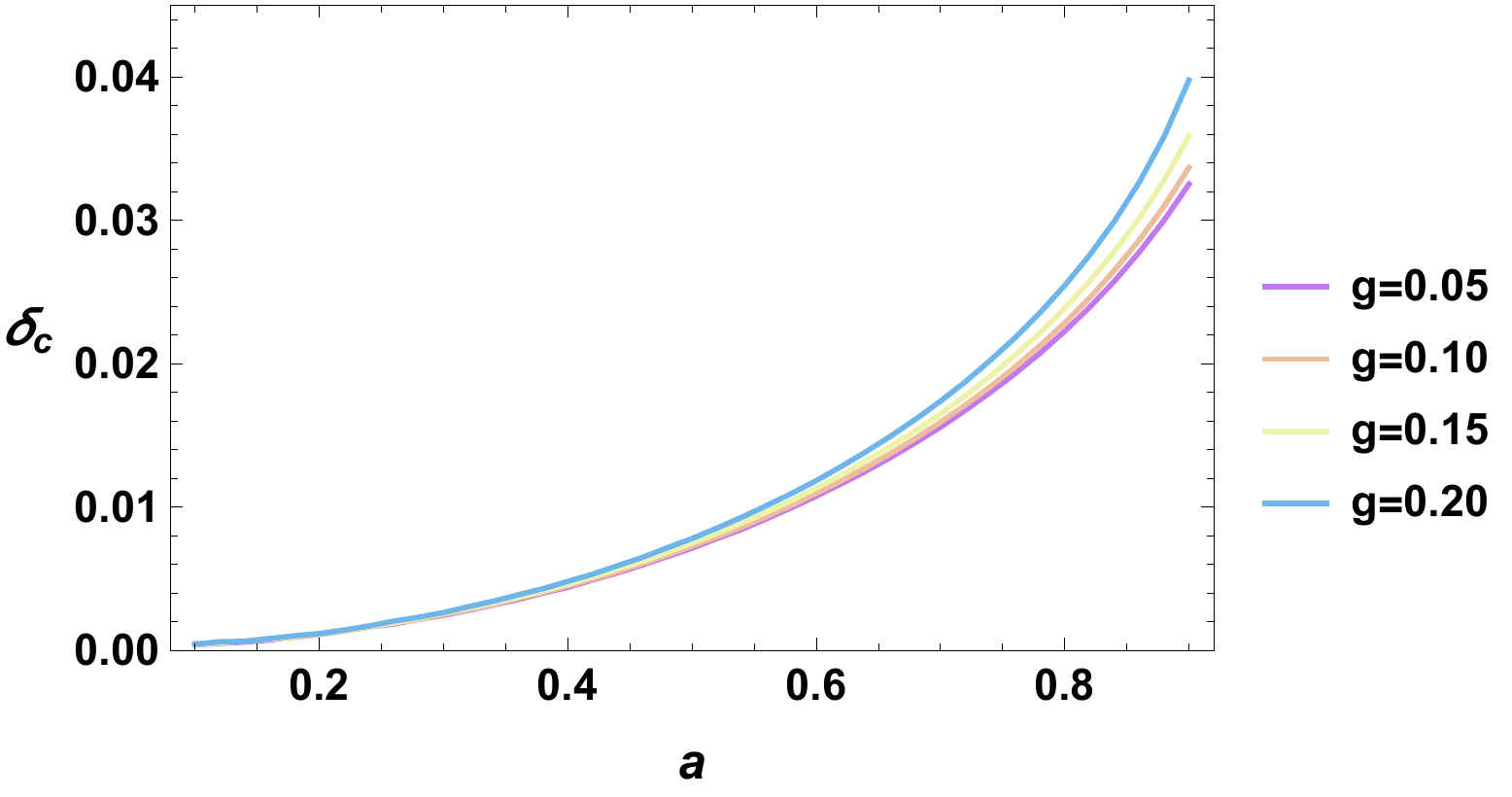} \\
		   \end{tabbing}
		    \vspace{- 1 cm}
\caption{{\it \footnotesize $R_c$ and $\delta_c$ parameters for rotating    Bardeen-AdS black holes  in terms of the ($a$, $g$) reduced moduli space}.}
\label{sh1}
\end{center}
\end{figure}

To   analysis th distortion  of the black hole shadows, 
 certain   deformation parameters, denoted by   $R_c$ and $\delta_c$, are needed.  These parameters     control the associated  size and  the shape, respectively \cite{hioki2009measurement,amir2016shapes}.   Concretely, the size is  configured   via three specific points. Indeed,   two of them are   the  top and  the bottom positions  of shadow ($x_t,y_t$) and ($x_b,y_b$). The third one concerns     the  reference circle  point denoted by ($\tilde{x}_p,0$).   In this geometric representation,  the point of  the distorted shadow circle ($x_p,0$) intersects the horizontal axis  at $x_p$. In this way, the  distance between these     points   is   given by  a parameter defined as follows $D_c= \tilde{x}_p- {x}_p=2R_c-(x_r-x_p)$ \cite{amir2016shapes}.   In certain approximations, $R_c$    takes the following  form 
\begin{equation}
\label{d1}
R_c=\frac{(x_t-x_r)^2+y_t^2}{2|x_t-x_r|}.
\end{equation}
In these optical behaviors,  the   distortion parameter   controlled by  ratio  of  $D_c$ and $R_c$   reads as 
\begin{equation}
\label{ dddd}
\delta_c=\frac{|D_c|}{R_c}.
\end{equation}
To go beyond such an optical study of  the rotating Bardeen AdS  black holes, we examine  the above  astronomical  quantities.   These parameters controlling the size and shape deformations are  illustrated  in Fig.(\ref{sh1}) using a  reduced moduli space coordinated by ($a$, $g$).
For small  values of  $a$ and $g$, we remark that  the shadow  parameter $R_c$ controlling the size  is large  contrary to  large values of $a$ and $g$. For fixed large values of $a$, the size parameter is decreasing by increasing $g$.  This behavior appears in the regular Bardeen black hole\cite{He,H2}. This can be linked to the circle geometry changing toward D-shape configurations by increasing $a$. For fixed large values of $g$, the size parameter is decreasing by increasing $a$.  \\  We move now to analyze the   distortion parameter  variation $\delta_c$. For fixed values of $a$, the distortion parameter  is increasing with $g$.  For small values of $a$, however,  this parameter remains small and constant.  Fixing   $g$, the geometric  parameter $\delta_c$  increases  with the rotating parameter and takes  small values for small values of $a$.  The obtained  results confirm the impact of $g$   on  the size and the  shadow form  observations.
\section{ Shadows of quintessential rotating Bardeen AdS black holes}
\begin{figure}[ht!]
		\begin{center}
		\centering
			\begin{tabbing}
			\centering
			\hspace{9.2cm}\=\kill
			\includegraphics[scale=.43]{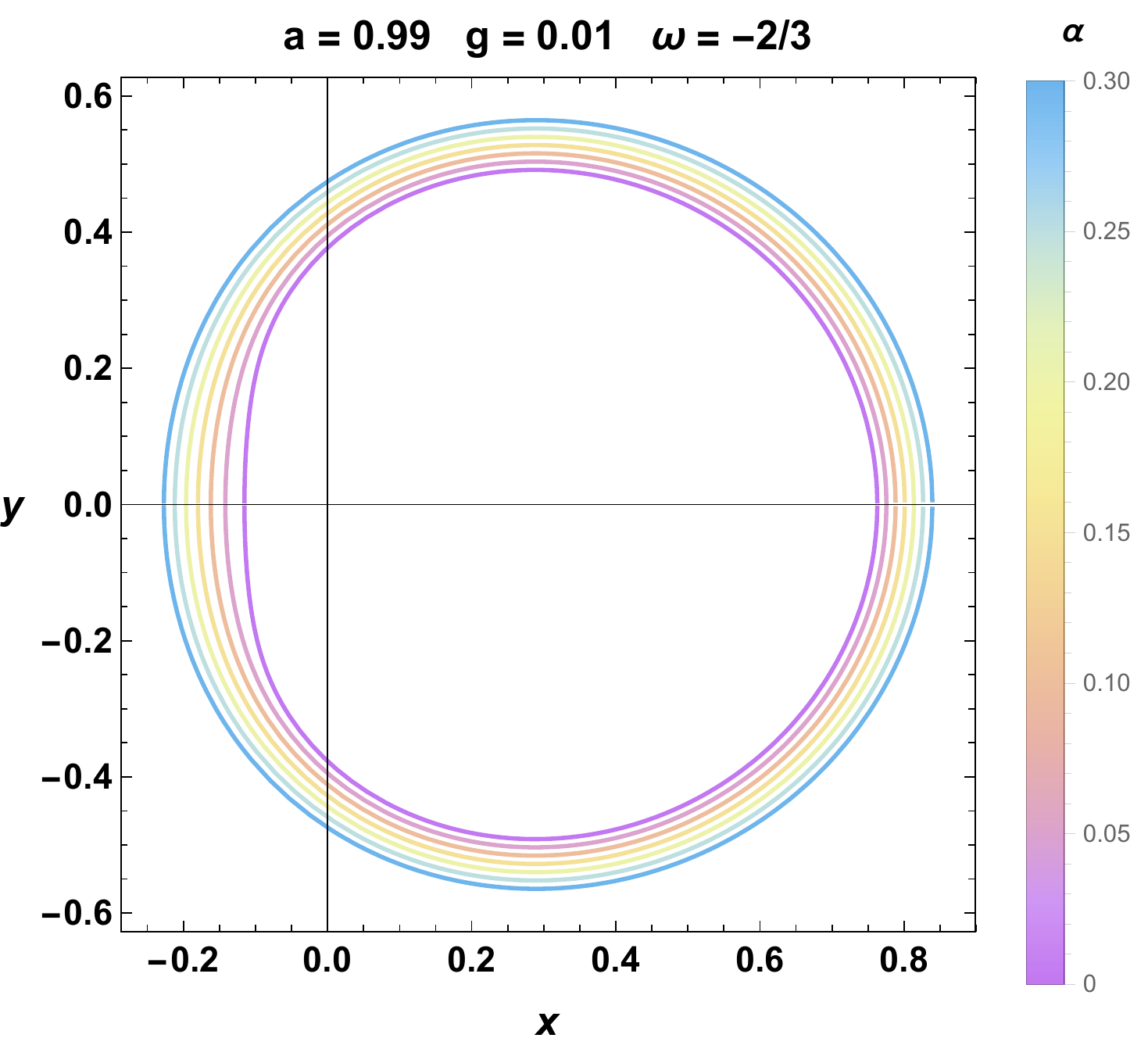} \>
			\includegraphics[scale=.43]{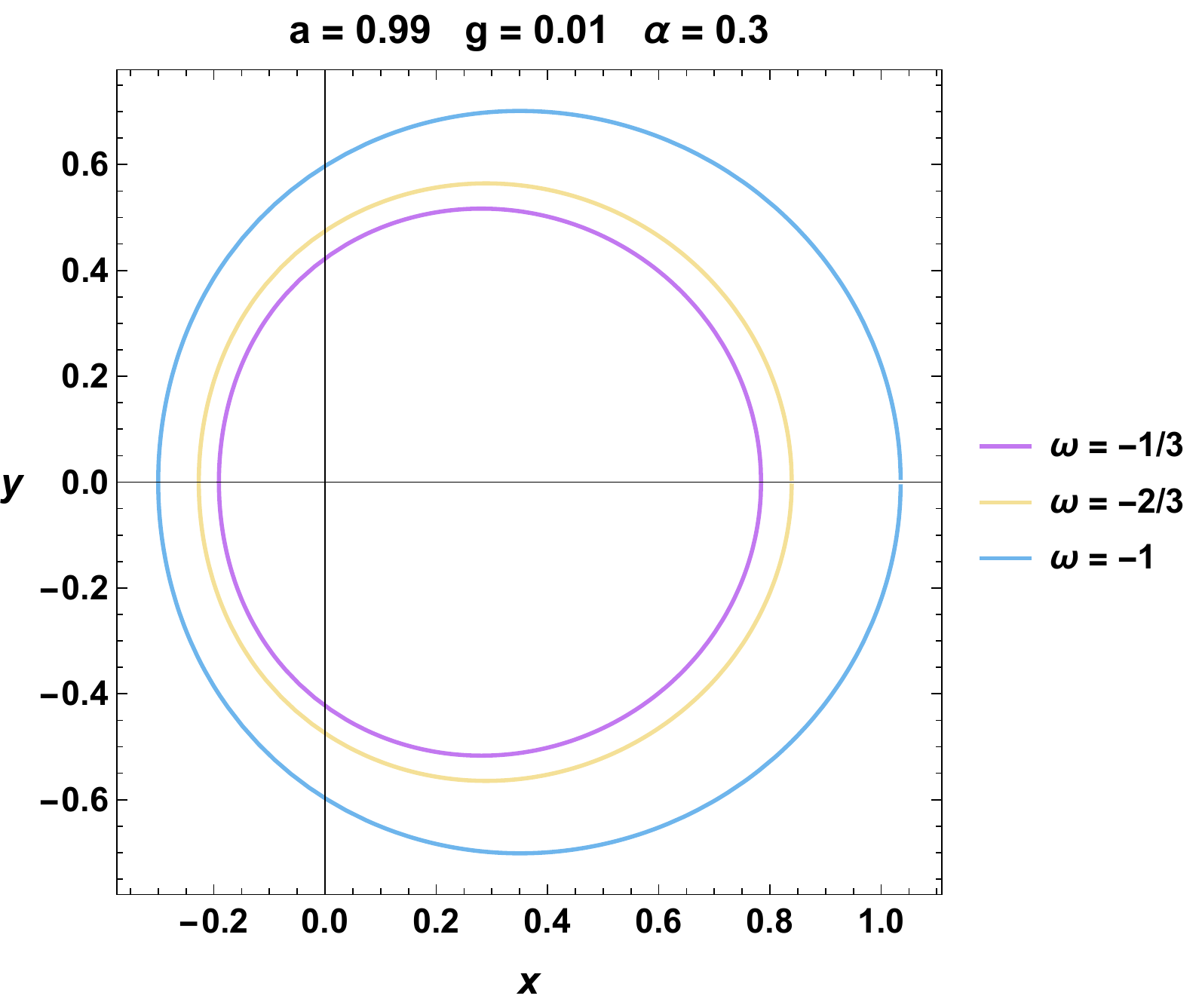} \\
			\includegraphics[scale=.43]{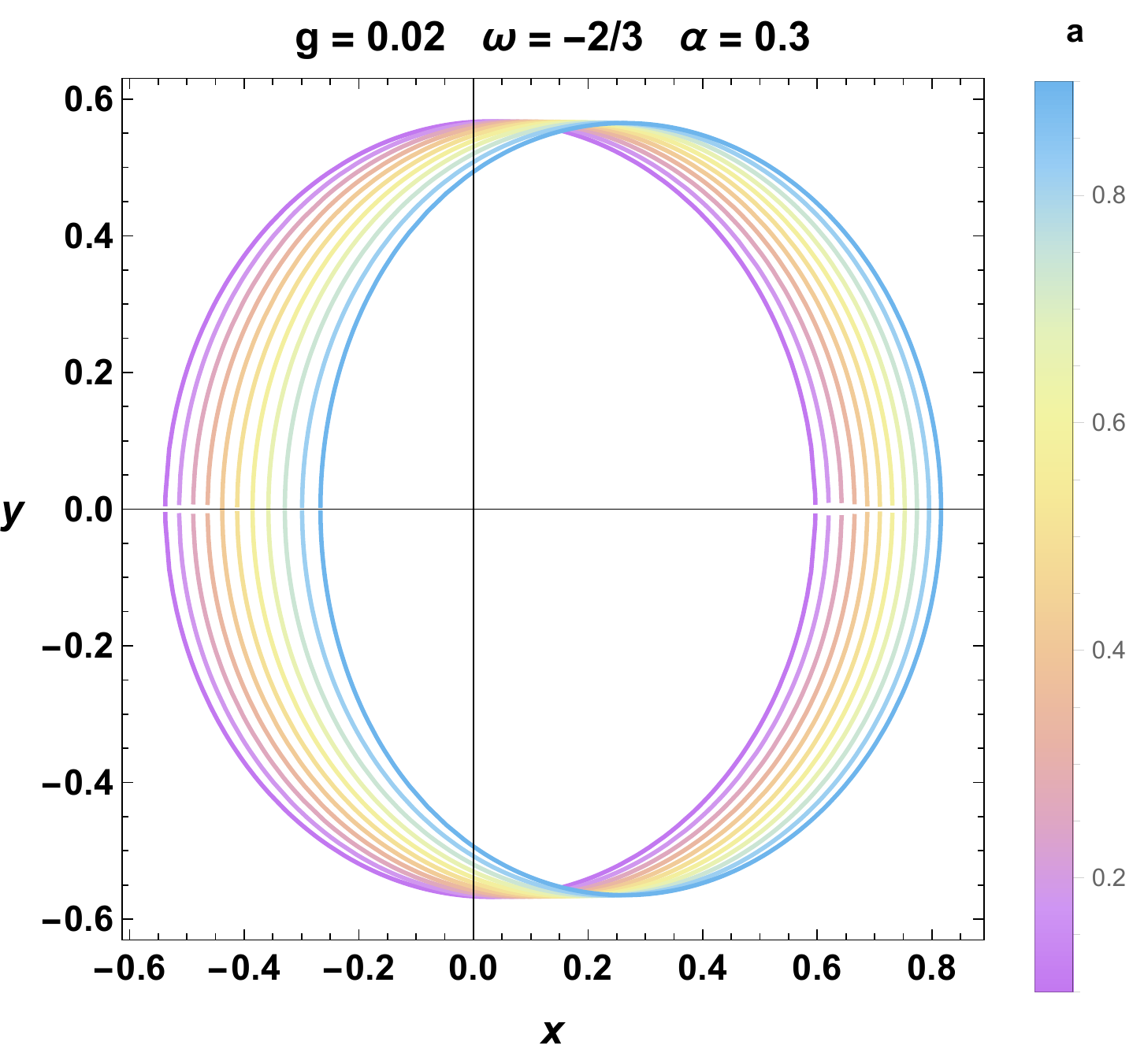} \>
			\includegraphics[scale=.43]{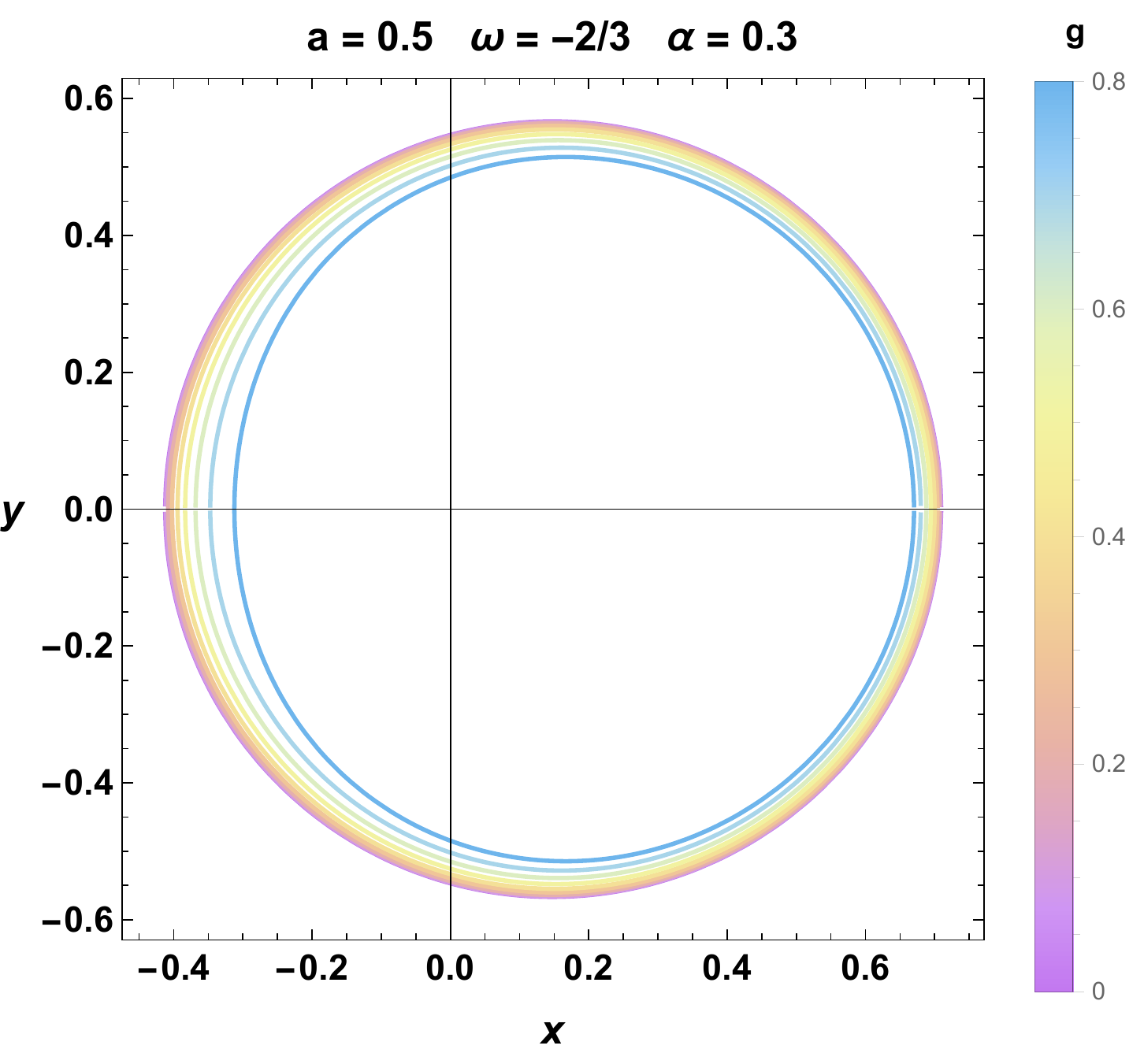} \\
		   \end{tabbing}
		  \vspace{-1 cm}
\caption{{\it \footnotesize  Optical shadow behaviors of  quintessential rotating    Bardeen-AdS black holes  for different values of  $a$, $g$, $\alpha$ and $\omega$}.}
\label{sh2}
\end{center}
\end{figure}

\begin{figure}[ht!]
		\begin{center}
		\centering
			\begin{tabbing}
			\centering
			\hspace{9.4cm}\=\kill
			\includegraphics[scale=.45]{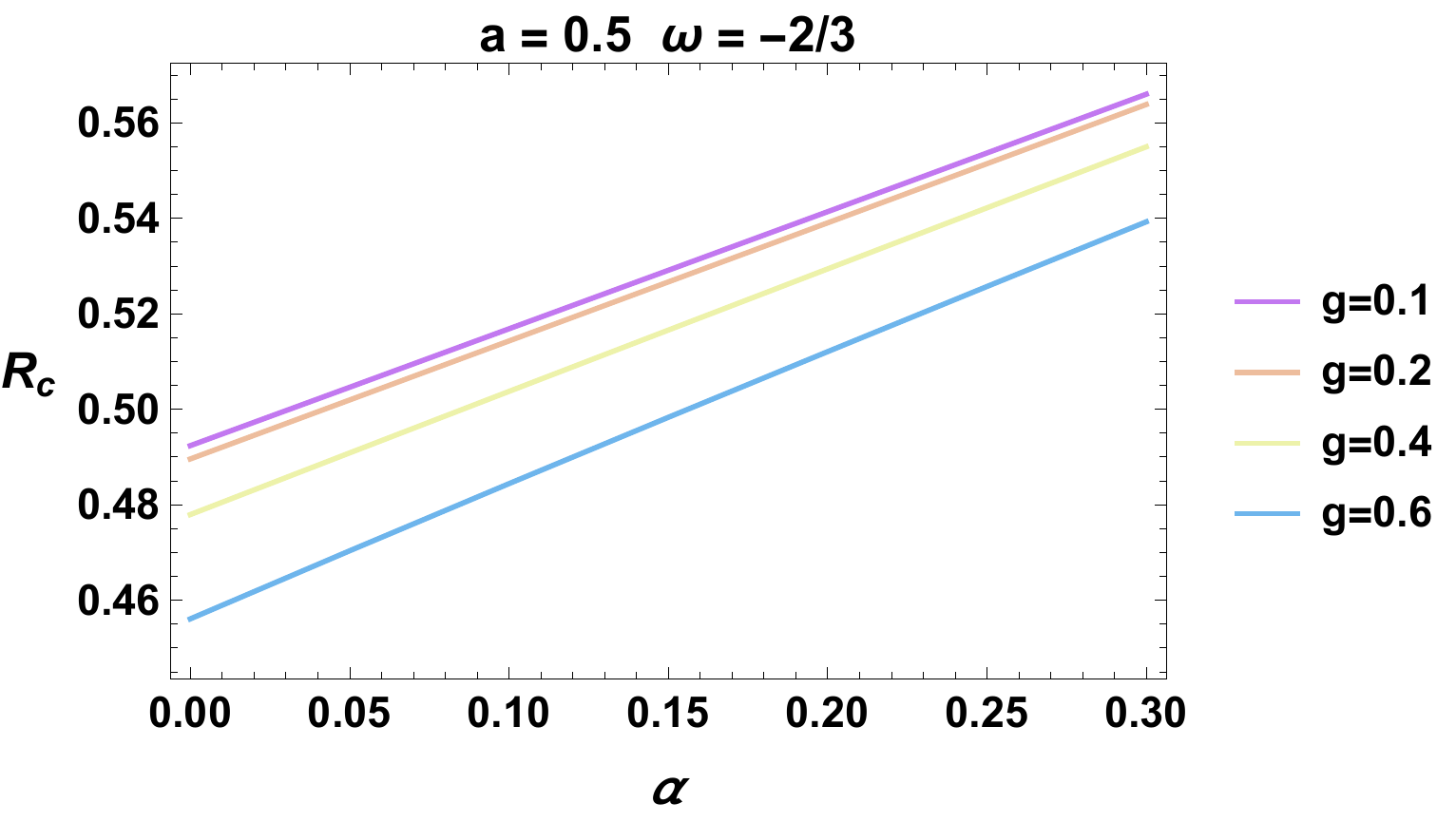} \>
			\includegraphics[scale=.45]{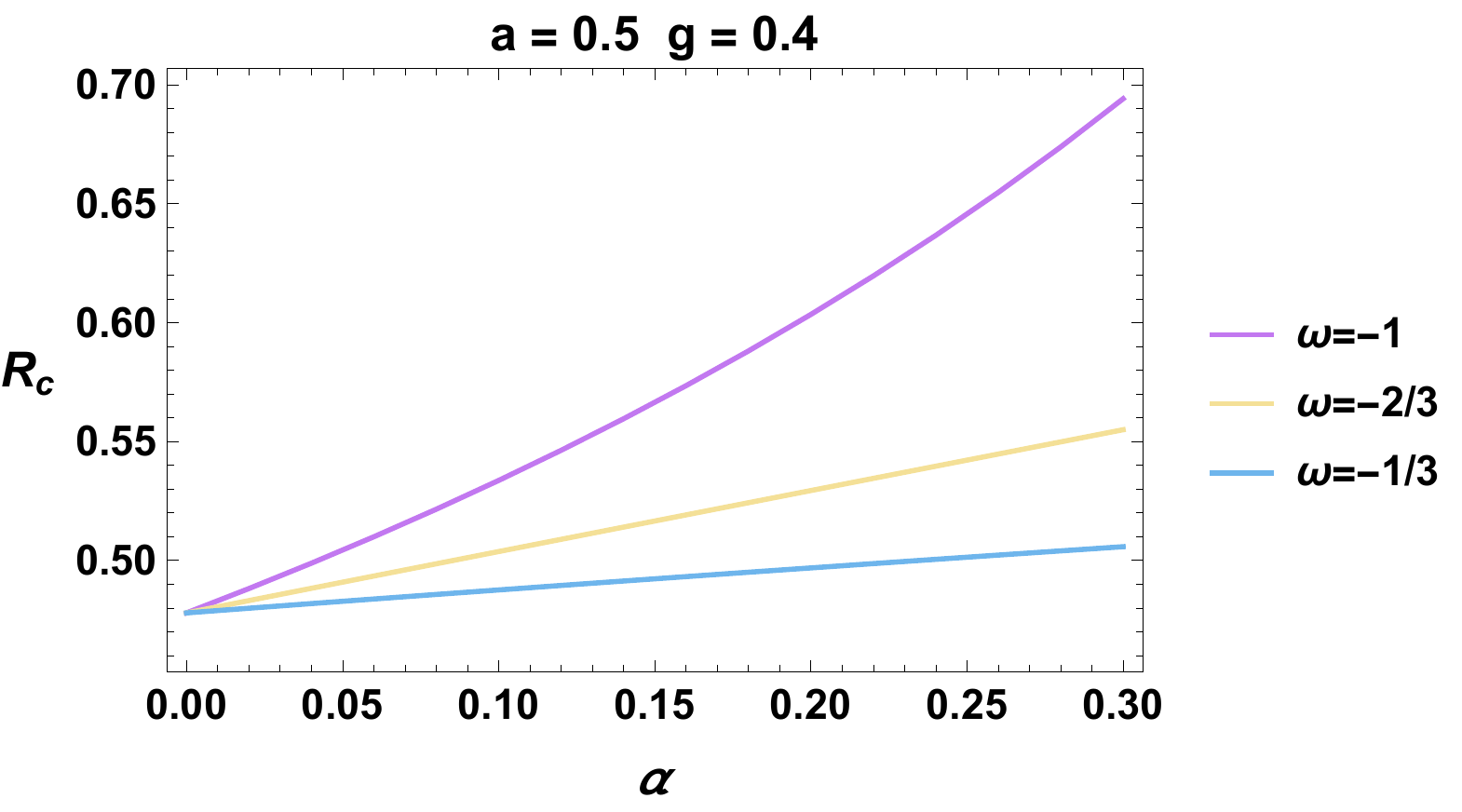} \\
			\includegraphics[scale=.45]{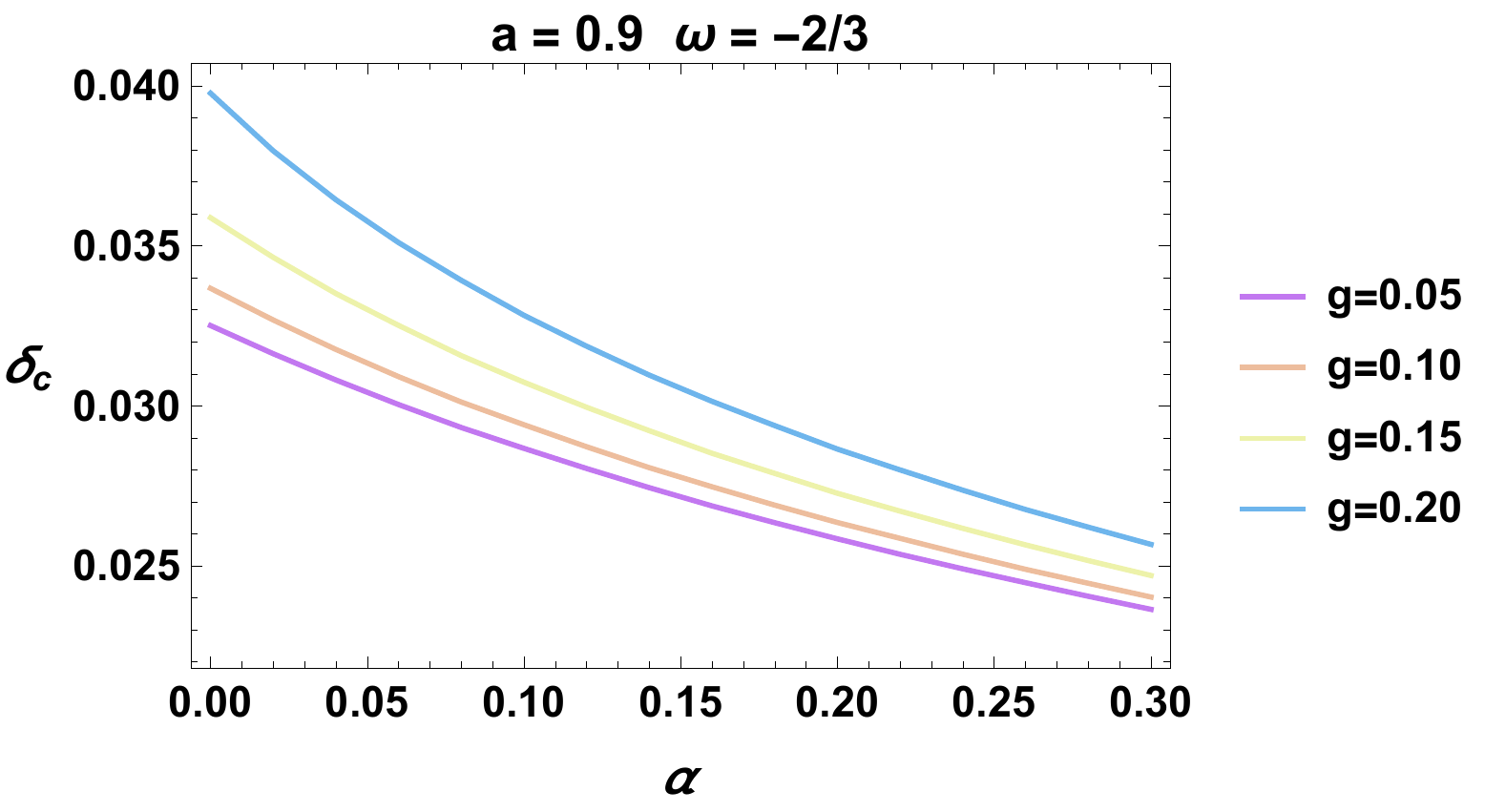} \>
			\includegraphics[scale=.45]{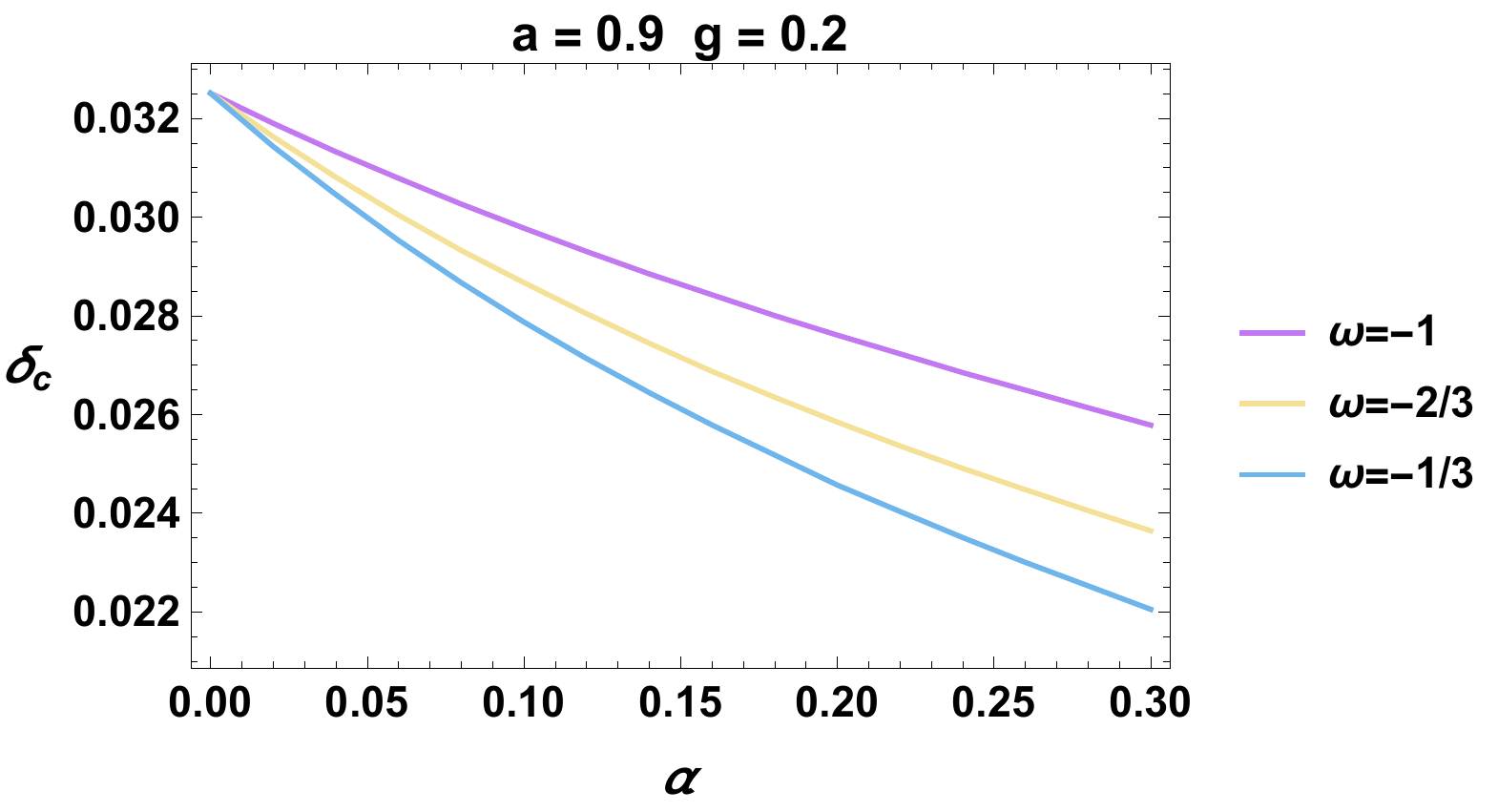} \\
		   \end{tabbing}
		   \vspace{- 1 cm}
\caption{{\it \footnotesize Behaviors of  $R_c$ and $\delta_c$ parameters for rotating   quintessential   Bardeen-AdS black holes  in terms of  dark sector parameters}.}
\label{qdis}
\end{center}
\end{figure}

In this section, we examine the shadow behaviors of  the quintessential rotating Bardeen AdS black holes. The computations will involve two extra parameters carrying information on dark field contributions\cite{x74}. In this way, $\Delta_{r}^\omega$ will be implemented in the previous calculations of the equations of motion. Many models could be dealt with by fixing the state equation parameter $\omega$. However, we consider only known values being investigated in different black hole backgrounds. In particular, we analyze three $\omega$-model shaving $\omega=-1, -1/3,  -2/3$. Roughly, the shadow aspects will be approached by varying the internal and external black hole parameters. These parameters control the associated  moduli space which can be decomposed as  follows
\begin{equation}
\label{ }
\mathcal{M}=\mathcal{M}_{int}\times\mathcal{M}_{ext}
\end{equation} 
where $\mathcal{M}_{ext}$ denotes the dark field sector.  As done in the previous section, one-dimensional shadow configurations could be illustrated by varying one parameter and fixing the remaining ones. The associated  shadow geometries are plotted in Fig.(\ref{sh2}).
Concerning the $\alpha$ variation, we examine two regions. For small values of this parameter, the size of the shadow decreases, and the D-shape appears. For large values of $\alpha$, however, the D-shape disappears and the size increases. Such behaviors are observed  in all $\omega$ models. Indeed,  the size increases by decreasing  $\omega$.   The  D-shape geometry disappears by taking $\alpha=0.3$. These geometrical configurations are not modified even we vary the rotation parameter $a$. Increasing this parameter, the circles are deviated from the origin. For $\alpha=0.3$, it has been remarked that the parameter $g$ controls the shadow size. Indeed, it decreases by increasing $g$. A close inspection shows that such a parameter controls the shape deformations contrary to the electric charge. It can play a similar role as the rotating parameter appearing in many black hole shadow activities. The appearance of the D-shape geometry is linked to the existence of three parameters, $a$  which controls the shape deformation in the ordinary black hole solutions,    $g$ and  $\alpha$.\\
As in the previous section,  we  analyze the associated distorted geometries by introducing the two extra parameters associated with the dark sector.   In particular,  we inspect the corresponding effect on $R_c$ and $\delta_c$ parameters.   By varying the  dark sector parameters and the charge of the nonlinear electrodynamics, these behaviors are illustrated in Fig(\ref{qdis}). 
Fixing  $a$ and $\omega$,   it follows from this figure that $R_c$ increases linearly  with $\alpha$ for different values of $g$.  It decreases by increasing $g$.   The later behavior appears for $\omega$ variations.   Contrary to   $R_c$,  $\delta_c$   decreases  by increasing  $\alpha$ for different values of $g$.  It increases with  $g$.    An inverse behavior appears for $\omega$ variations. For fixed  values of $\omega$,  the geometric parameter $\delta_c$ decreases by increasing $\alpha$.  The obtained results confirm the impact of $g$ and $\alpha$  on the size and the shadow shape observations.

    \section{ On observations in the light of  the $M87^\star$  picture}
\begin{figure}[!ht]
		\centering
			\begin{tabbing}
			\hspace{6.cm}\= \hspace{6.cm}\=\kill
			\includegraphics[scale=.438]{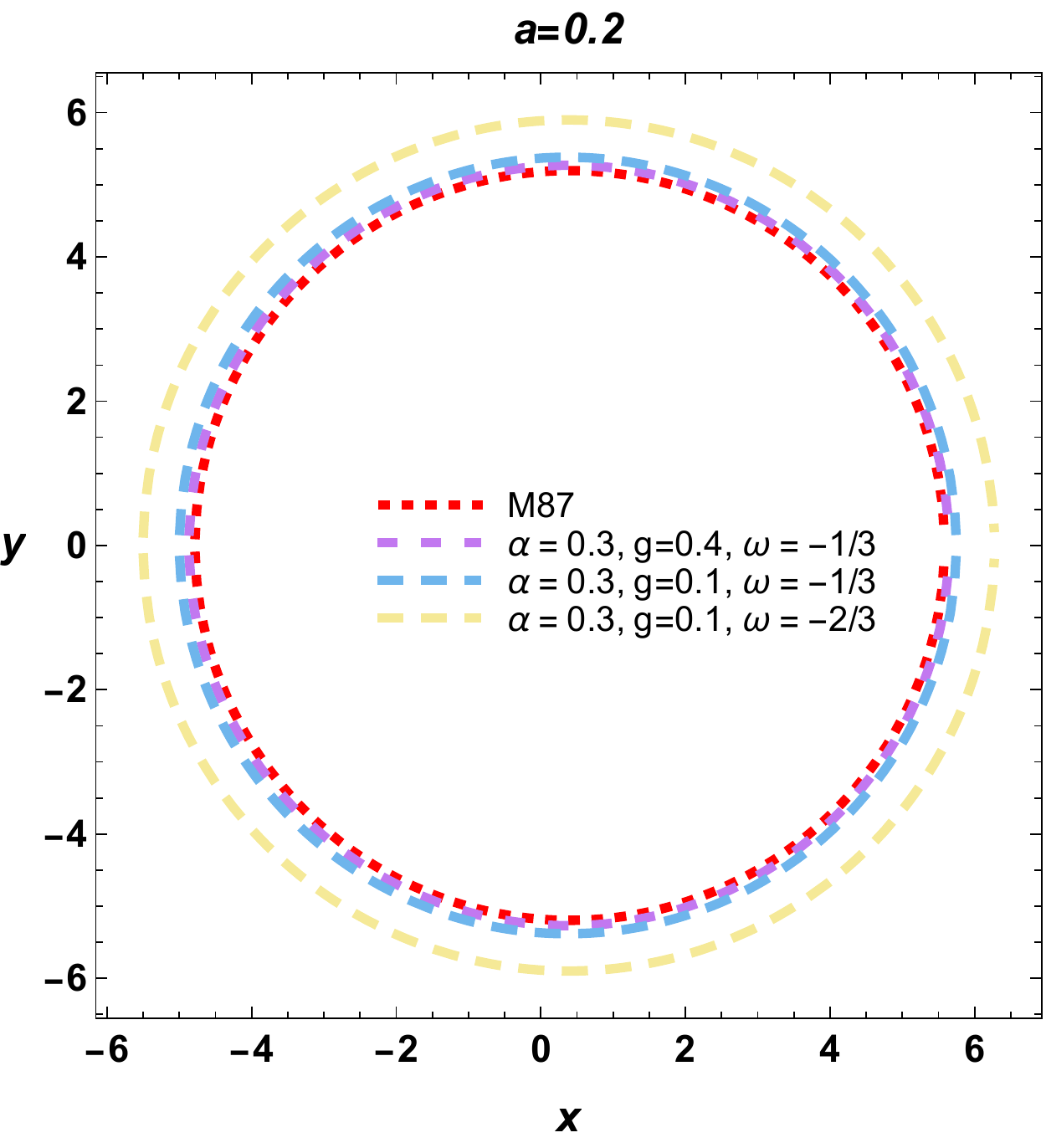} \>
			\includegraphics[scale=.438]{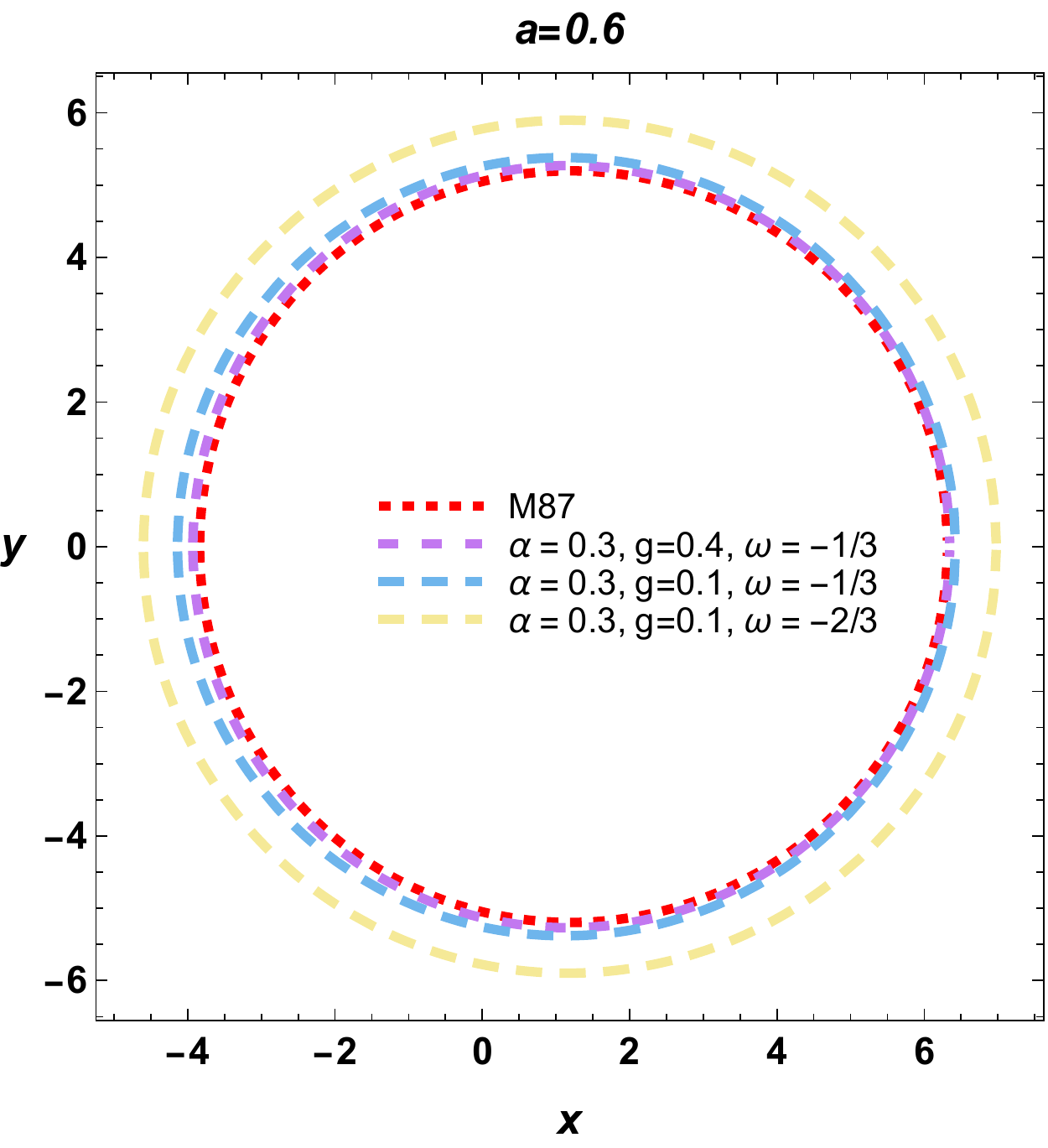} \>
			\includegraphics[scale=.438]{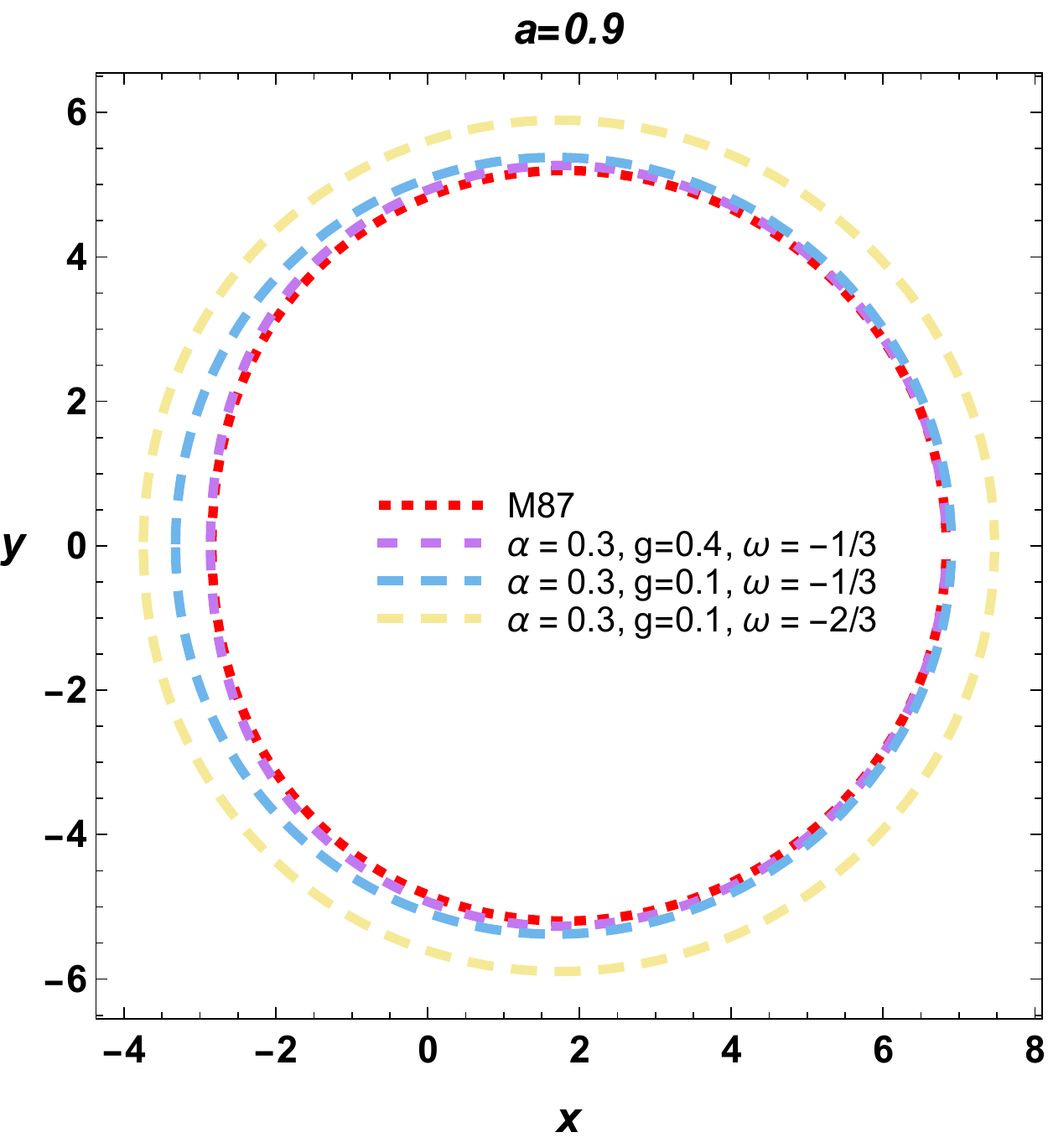} \\
		          \end{tabbing}
		          \vspace{-1cm}
\caption{\it \footnotesize  { Black hole shadows for different values of  the rotating parameter $a$,    compared  to the M87$^*$  shadow, using  $\lambda= -0,002$   and  $M = 1$  in units of the M87 black hole mass given by $M_{BH} =6.5\times 10^{9}M_\odot$ and $r_0 =91.2 kpc$.}}
\label{f222}
\end{figure}

A close examination shows that the recent observational results corresponding the shadows of the supermassive black hole  $M87^\star$,    realized by  EHT international collaborations,  have  provided many promoting approach to probe
certain gravity theories  and alternative modified physical   models.  Motivated by such activities,   it  would  be interesting to make contact with such    observational fundings.  It  has been remarked   that the    observational data  could  put   constraints on the black hole  moduli space\cite{3,MA1,R2,R3}.   Precisely, the shadow behaviors can be   considered in terms of the external sector of the moduli space for different values of the rotation parameter $a$.     In the unit of the M87$^\star$ mass, we could superpose the M87$^\star$ black hole shadows and the one of the present work by considering appropriate black hole parameters.  These parameters could provide contact with such observational findings. Considering  different values of  $a$,
the  associated behaviors are shown in Fig.{\ref{f222}}.
Comparing  the   shadow radius of the both black holes illustrated  in  this figure,  we reveal   that the M87$^*$ shadow  coincide perfectly with  such  rotating  quintessential AdS black holes    for  specific values of the  relevant   parameters.    Varying the rotating parameter, the computations  show that the relevant parameters are   $\alpha= 0.3$,   $g= 0.4$,  and   $\omega= -1/3$.   For $\omega=-2/3$,  we notice that  the shadow of   the rotating  quintessential AdS black holes  can    approach  the M87$^*$ shadow.  For non distorted geometries,  it  has been remarked similar behaviors like the M87$^*$.    Since  for $a=0.9$ the shadows have been distorted, it could be interesting to   discuss    the distortion of the shadow compared by  M87$^*$  one.    It has been observed    that the distortion of the  quintessential AdS black hole is  the same as M87$^*$ for $\alpha= 0.3$,   $g= 0.4$,  and   $\omega= -1/3$. For other value parameters,  we  remark  that the distortion is different that   the   M87$^*$ one.  This graphic   analysis  could confirm the  above specific constraints  on  such black hole parameters.

\section{Conclusion}
In the  present  work,  we have analyzed  the  shadow geometrical aspects of rotating Bardeen black holes in different backgrounds including AdS geometries and dark sectors. First, we  have considered rotating  Bardeen AdS black  hole solutions without external dark fields.   As the rotating parameter, we have shown that    the charge of the nonlinear electrodynamics affects   not only the space-time structure of the black hole, but also  controls the  shape deformation.   It has been remarked that it also affects the size. Then, we have discussed  the geometrical observables describing such shadow geometric deformations.   Implementing dark sectors, we have investigated  shadow properties  of the quintessential rotating Bardeen AdS black holes. In particular, we have inspected the effect of quintessential dark field on the shadow curves showing interesting features.    Increasing the dark fiend intensity,  the shadow size increases, and the D-shape geometry disappears producing  the circular perfect    geometries.  In this way, we have observed that DE  affects only the size but also deforms the shadow shape.    It has been shown that the appearance of the D-shape geometry has been linked to the existence of three parameters  being $a$, controlling the deformation shape in the ordinary black hole solutions,  $g$ and $\alpha$.    It has been remarked that  the parameter $\alpha$ associated with the  quintessence field  not only plays  a thermodynamic  important role  but  appears  also  as a  relevant geometric  parameter which can  control the  black hole optical aspects.   Precisely,  it controls the size and the shape of the shadow through the existence of other parameters $g$ and $a$, being related to  the space-time configurations  and the  rotating parameter of black hole,  respectively.  Finally, we   have  provided   a possible link 
with observations from Event Horizon Telescope    by  revealing   certain constraints on the involved  black hole parameters in
the light of the M87$*$ picture. \\
Many open questions could   be addressed.  It would be interesting to   consider  the  dark matter  influence  as a relevant element in  dark sector contributions.   In particular,  we could inspect  optical  modifications which could come from such a matter. Higher dimensional  solutions could be  also a possible issue. We hope to come back to these questions in future works.
\section*{Acknowledgements}

The authors  would like to thank  Y. Hassouni for discussions and collaborations and they would like
also to thank the anonymous referee for interesting comments and
suggestions.     This work is partially supported by the ICTP through AF.

\end{document}